# Job market effects of COVID-19 on urban Ukrainian households

Tymofii Brik, Kyiv School of Economics, Kyiv, Ukraine
https://orcid.org/0000-0002-5542-1019

Maksym Obrizan,[1] Kyiv School of Economics (KSE), and Research Consulting & Development, Kyiv, Ukraine
http://orcid.org/0000-0002-0924-0671

**Abstract**
The employment status of billions of people has been affected by the COVID epidemic around the Globe. New evidence is needed on how to mitigate the job market crisis, but there exists only a handful of studies mostly focusing on developed countries. We fill in this gap in the literature by using novel data from Ukraine, a transition country in Eastern Europe, which enacted strict quarantine policies early on. We model four binary outcomes to identify respondents (i) who are not working during quarantine, (ii) those who are more likely to work from home, (iii) respondents who are afraid of losing a job, and, finally, (iv) survey participants who have savings for 1 month or less if quarantine is further extended. Our findings suggest that respondents employed in public administration, programming and IT, as well as highly qualified specialists, were more likely to secure their jobs during the quarantine. Females, better educated respondents, and those who lived in Kyiv were more likely to work remotely. Working in the public sector also made people more confident about their future employment perspectives. Although our findings are limited to urban households only, they provide important early evidence on the correlates of job market outcomes, expectations, and financial security, indicating potential deterioration of socio-economic inequalities.

***JEL Codes*** *J64, D14, D84*

---

[1] Correspondence to mobrizan@kse.org.ua

# 1 Introduction

No country has escaped the pandemic crisis of COVID-19. According to the estimations of the World Bank, the epidemic is pushing from 40 to 60 million people globally into extreme poverty (Mahler et al., 2020). In light of this crisis, there is a salient demand for robust data on the economic activities of individuals and households. These data are essential for two reasons. First, household expectations are one of the key factors that influence the economy (Baldwin and Mauro, 2020). Second, household-level analysis is particularly important for planning social policies aimed to protect vulnerable groups, reduce inequalities, and ensure the well-being of citizens (Deaton, 2005).

At the same time, such data about households during the COVID-19 crisis are lacking. Currently, most of the recent studies (including working papers) primarily focus on macro-economic consequences of the COVID-19 crisis (e.g., Barro, Ursua and Weng, 2020; Eichenbaum, Rebelo and Trabandt, 2020; Fernandes, 2020). Those rare ongoing studies about households' consumption and expectations have covered primarily the US (Baker et al., 2020; Coibion, Godonichenko and Weber, 2020) and OECD countries (Ambrocio, 2020; Rothwell and Van Drie, 2020) omitting other parts of the world. Our goal is to close this gap and shed light on how households respond to the COVID-19 crisis in a transition country. We present new data about Ukraine in order to address this gap and contribute to the fast-growing literature on the COVID-19 and household economies.

In this paper, we focus our attention on the job market effects of the epidemic on urban households in Ukraine. We model which respondents are likely to lose jobs; to work from home due to the COVID-19; to fear losing a job; to have savings for less than one month if the quarantine continues. Our data allow us to include a vast array of socio-demographic and socio-economic factors describing respondents and household status. We control for gender, age, marital status, children, education, financial status, industry, and occupation. We observe that the crisis hit harder more vulnerable and less resourceful Ukrainians. These findings indicate that the crisis is likely to exacerbate socio-economic inequalities: wealthier and better educated Ukrainians in large cities are more likely to secure their jobs, whereas vulnerable Ukrainians are likely to be pushed back even further.

# 2 Ukrainian context

## 2.1. Economic background

Ukraine is a transition country in Eastern Europe, which starting from 1991, has witnessed a major shift from the planned economy of the authoritarian Soviet regime to the market economy and democratic institutions. Considering the region, Ukraine has been regarded as one of the least successful transition countries in terms of GDP and wages (Braithwaite, Grootaert and Milanovic, 2016; Guriev, 2018). The GDP of Ukraine collapsed by half from 1990 to 1994 with a slow decline between 1994 and 2000 (Sutela, 2012). On average, transition countries increased per capita incomes by around 50% of what they were in 1989, whereas Ukraine did not achieve such outcomes (Guriev, 2018). Moreover, while in other transition countries, at least some income groups managed to achieve salaries similar to the G7 countries, in Ukraine this has not been the case. The situation has somewhat improved between 2001 and 2008 but mostly due to better terms of trade (higher prices of major exports such as metals and lower prices of Russian gas) rather than reforms.

Ukrainian economic system has been often described as oligarchic and rent-seeking (Gorodnichenko and Grygorenko, 2008). Such an institutional environment has long prevented

Ukraine from stable economic growth. Researchers often explain slow Ukrainian development as the result of the weak rule of law, closed ties between political power and economic elites, and corruption (Guriev, 2018; Milanovic, 1998; Restrepo et al., 2015; Sutela, 2012; World Bank, 2019). At the same time, the population decline from 1989 till the early 2000s was more salient in Ukraine than in neighboring countries. Ayhan, Gatskova and Lehmann (2020) point out that Ukraine lost about nine million people from 1991 to 2016. They attribute this loss to lower fertility rates, high mortality, and out-migration.

In terms of economic expectations and attitudes, Ukrainians tend to have gloomy views about their economy. For instance, they tend to significantly overestimate the rates of economic inequality in the country (Gimpelson and Treisman, 2018) and be quite skeptical of market reforms. In 1992, in the wake of independence, 64% of Ukrainians said that they shared rather positive attitudes towards land privatization. In 2018, 25 years after, only 20% thought the same (Brik and Shestakovskyi, 2020). Researchers explain this phenomenon with a difficult transition period, weak governmental institutions, poor local governance (lack of decentralization), and underdeveloped local economic activities (lack of participatory budgets, lack of land market reforms, the strong influence of clientelism).

Since 2014 the Ukrainian economy has been further damaged by the annexation of Crimea and the hybrid war with Russia. Losing parts of the Donetsk and Luhansk regions that specialized in mining industries were particularly harmful to the economy. This industry accounted for 25% of total exports and 15% of the total GDP of Ukraine (Havlik 2014). The destruction of physical capital, mass migration, and market disintegration after the war contributed to the diminishing of Ukrainian GDP (Coupe and Obrizan, 2016). Some previous studies of the individual-level Labor Force Survey during 2004-2013 showed evidence of "job polarization in Ukraine, with relative increase employment of managers, professionals, services and sales jobs, and elementary occupations, and with a significant decline in skilled manual occupations, clerks and technicians" (Kupets, 2016 p. 25). The authors concluded that Ukraine has a significant mismatch between education and labor markets and that almost 40% of employed respondents were overeducated. According to the Ptoukha Institute for Demography (2019), the percentage of people who live below the subsistence level (i.e., absolute poverty) was 43,2% in 2018. According to the same data, children below 17 and retired people (especially women) were most vulnerable to poverty. Considering the household structure, the most vulnerable to poverty was the household with three and more children and those households with at least one unemployed.

**2.2 COVID-19 in Ukraine**

According to the OECD report (2020), there are about 57 thousand confirmed cases in Ukraine by July 2020. Among them, 1,456 resulted in death, and 29,769 were recovered. These are moderately low numbers given the size of the Ukrainian population.[2] Although the government was relatively slow with testing, coronavirus containment policies were implemented quite rapidly, with just a few confirmed cases and not a single death. A three-week nationwide quarantine was initially imposed on March 12th, 2020, which shut down all educational institutions moving classes online.[3] Non-citizens were banned from entering the country on March 13th, and all national and international air and rail travel was banned on March 17th. A mandatory hospital observation or self-isolation for 14 days was required for everyone entering

---

[2] Ukraine ranks 87th in terms of total cases per 1 million people and 78th in terms of deaths per 1 million people: https://www.worldometers.info/coronavirus/?utm_campaign=homeAdvegas1?%22%20%5Cl%20%22countries#countries
[3] http://www.golos.com.ua/article/328891

Ukraine.[4] Wearing masks or respirators in public places became obligatory with considerable fines for violation in the range from 17,000 to 34,000 UAH, which was about 700-1,500 USD.[5] These bans were relaxed only in mid-June.

Some early surveys showed that urban Ukrainians varied in compliance with quarantine. For example, a survey conducted on March 15th (Liga, 2020) showed that women tend to wash their hands more often than men (91% vs. 78%), and they are also more likely to use sanitizers (57% vs. 41%). Furthermore, most Ukrainians thought that their chances of getting infected with COVID-19 were not very high (mean value of 4.5 out of 10). The same survey showed that 80% of employed urban Ukrainians were eager to continue going to their workplaces despite the threat of the virus. To the best of our knowledge, there has been no systematic attempt to investigate household economic expectations and job prospects in Ukraine during the COVID-19 pandemic. In what follows, we present our statistical analyses of a unique dataset to address this question.

**3 Data**

**3.1 Gradus survey**

We use the survey conducted by the Ukrainian research firm Gradus, which developed a smartphone application to recruit respondents and circulate questionnaires. Initially, all respondents were recruited from the general population of urban Ukrainians using a variety of methods, including probability-based sampling, face to face and phone interviews, distribution of promo codes, online social media advertising. The panel excludes those who live in the conflict zones in Ukraine's east as well as the Crimean Peninsula that are currently outside its government's control. Gradus applies weights based on gender, age, size of the settlement, and macro-regions to make the panel equivalent to the Ukrainian urban adult population under age 60. Hence, the trends discussed below may not generalize to citizens who are older or located in rural areas. Respondents receive questionnaires on different subjects approximately every week. Respondents receive questionnaires on different subjects approximately every week.
We used the survey conducted on April 8th, which asked about the Orthodox Easter celebration, compliance with the stay-at-home policies, and employment changes due to the COVID-19. The survey lasted less than 24 hours, and 1,176 respondents out of 2,177 listed in the Gradus panel have responded (implying the response rate of 54%).

The main caveat of the data is the exclusion of rural territories, which clearly limits our understanding of labor markets in the countryside and smaller cities with a population of less than 50 thousand. Nevertheless, urban Ukraine is quite diverse geographically and in terms of skills and industries (Kupets, 2016; Ayhan, Gatskova and Lehmann, 2020). We focus our attention on those variables that can explain the change in the employment status in urban Ukraine and provide our insights for the respective policies. Our study can be extended once rural areas are added, but we suspect that economic activities in cities are more affected by quarantine given higher population density.

**3.2 Variables**

We constructed four *dependent variables* related to the job market effects on the household well-being:

---

[4] https://mfa.gov.ua/en/news/mfa-ukraine-q-coronavirus-covid-19-quarantine-measures-entering-ukraine-obtaining-consular-support
[5] https://www.kmu.gov.ua/en/news/rozyasnennya-shchodo-novih-obmezhuvalnih-zahodiv-na-period-karantinu

"Not working" – equal to 1 for respondents who answered that there is no work for them and they are fired, on paid or unpaid leave (and 0 otherwise)

"Working from home" – equal to 1 for those who work from home part or full time (and 0 otherwise)

"Fears to lose a job" – equal to 1 for respondents who are afraid to lose a job (and 0 otherwise)

"Savings for <1 month" – for respondents who have enough financial resources for one month or less (and 0 otherwise)

The first two variables are only asked of respondents who worked before quarantine, while the last two are asked of all respondents. However, for the dependent variable "Fears to lose job" we only include respondents who actually had a job before the quarantine.

For each of the dependent variables we have estimated the following linear probability model with robust standard errors clustered at the city level

$$DV_i = \alpha_0 + SD \cdot \alpha_{SD}' + RC \cdot \alpha_{RH}' + FS \cdot \alpha_{FS}' + IN \cdot \alpha_{IN}' + OC \cdot \alpha_{OC}' + \varepsilon_i, \quad (1)$$

where $\varepsilon_i$ is the individual error term for respondent $i$.

In these models we include a rich set of socio-demographic factors (SD), regional characteristics (RC), measures of financial status (FS), and indicators for industry (IN) and occupation (OC).

Specifically, socio-demographic factors (SD) control for the female gender, three age groups of 25-34, 35-44 and 45+ years old (with a base age group of 18-24 years old), indicators for married and never married, an indicator if there are children in the household, indicators for 2, 3 or 4+ members in the household, dummies for post-secondary and higher education, an indicator whether respondent knows the number of COVID cases within +/-10% of the actual number on the data and a dummy whether a respondent believes in God.

Regional characteristics (RC) include dummies for four regions of Ukraine and Kyiv (with the central region being the base), an indicator for living in a city with 100 thousand to 1 million inhabitants, and another one for living in a city with more than 1 million people (with cities smaller than 100 thousand people serving as a base). We account for financial status (FS) using indicators for owning a place to live, owning a car, and an indicator for not working before the quarantine (included in the last model for having enough savings for 1 month or less). There are also two indicators for self-assessed financial status: "Middle" if a household can afford food and clothes but not larger home appliances like a fridge, and "Wealthy" if a household can buy more expensive stuff (with those who cannot afford a mobile phone, clothes or even food being in a base category). A great advantage of our dataset is that we also have indicators for the industry or the sector of economic activity (IN) and occupation (OC) with descriptive statistics provided in Appendix Table 3.

## 4 Results

### 4.1 Descriptive Statistics

The descriptive statistics for the respondents are provided in Table 1. In the sample, 62% of respondents were working full time, 12% were working part-time, and 26% were unemployed. In total, 24.8% of respondents were not working because of being fired or taking paid or unpaid leave. A considerable fraction of respondents (40.4%) worked from home part or full time. 24.6% of respondents were afraid to lose jobs, including those who did not work before the quarantine. However, in our preferred sub-sample of 874 respondents who had jobs before COVID-19, there were 259 (or 29.6%) of such respondents. The most striking observation is that 56% of respondents did not have savings for more than 1 month if quarantine were to be continued.

Table 1. Means and standard deviations of socio-demographic and socio-economic characteristics

| Variables | # of Obs. | Mean | St. Dev. |
|---|---|---|---|
| Not working | 874 | 0.248 | 0.432 |
| Working from home | 874 | 0.404 | 0.491 |
| Fears to lose job | 1,176 | 0.246 | 0.431 |
| Savings for <1 month | 1,176 | 0.560 | 0.497 |
| Female | 1,176 | 0.599 | 0.490 |
| Age 25-34 | 1,176 | 0.355 | 0.479 |
| Age 35-44 | 1,176 | 0.292 | 0.455 |
| Age 45+ | 1,176 | 0.197 | 0.398 |
| Married | 1,176 | 0.646 | 0.478 |
| Never Married | 1,176 | 0.255 | 0.436 |
| Has children | 1,176 | 0.489 | 0.500 |
| 2 members in HH | 1,163 | 0.271 | 0.445 |
| 3 members in HH | 1,163 | 0.312 | 0.464 |
| 4+ members in HH | 1,163 | 0.329 | 0.470 |
| Post-secondary education | 1,176 | 0.091 | 0.288 |
| Higher education | 1,176 | 0.725 | 0.447 |
| Knows # of COVID cases | 1,160 | 0.234 | 0.423 |
| Believes in God | 1,176 | 0.629 | 0.483 |
| East | 1,176 | 0.120 | 0.325 |
| Kyiv | 1,176 | 0.241 | 0.428 |
| North | 1,176 | 0.115 | 0.319 |
| South | 1,176 | 0.136 | 0.343 |
| West | 1,176 | 0.164 | 0.371 |
| City 100-1M | 1,176 | 0.418 | 0.494 |
| City 1M+ | 1,176 | 0.448 | 0.498 |
| Owns house/apartment | 1,163 | 0.696 | 0.460 |
| Owns car | 1,163 | 0.427 | 0.495 |
| Middle financial status | 1,163 | 0.299 | 0.458 |
| Wealthy financial status | 1,163 | 0.362 | 0.481 |
| Did not work before quarantine | 1,176 | 0.257 | 0.437 |

*Note: Authors' calculations based on the Gradus survey.*

In our sample, there were 59.9% of females, 35.5% of respondents in the 25-34 age group, 29.2% were in the 35-44 age group, and 19.7% were age 45+ (with the remaining group being in the 18-24 age category). 64.6% were married, while 25.5% had never been married, and 48.9% of households had children. 27.1% of households had 2 members, 31.3% had 3 members, while 32.9% had 4 or more members with the remaining share being single-member households in a base category. 9.1% and 72.5% reported having post-secondary and higher education correspondingly. The high share of people with higher education can be partially explained by the sample composition and partially by a lower quality of education and a large number of "diploma mills" in Ukraine (Obrizan 2019). Only 23.4% of respondents knew the number of reported COVID cases on the date of interview in the world within +/-10%.

24.1% of respondents lived in Kyiv, while 11.5% to 16.4% of participants lived in other regions. 41.8% lived in cities with a population from 100 thousand to 1 million, and 44.8% lived in cities with more than 1 million inhabitants. 69.6% owned their place of living and 42.7% had a car. 29.9% and 36.2% reported being in the "Middle" and "Wealthy" self-rated financial states, and 25.7% did not work before the quarantine.

**4.2 Regression results**

We estimated linear probability models (LPMs) for our four dependent variables related to the job market and financial prospects of urban Ukrainian households. Despite certain limitations, like the possibility of obtaining predicted probabilities outside of 0 to 1 range, LPM has a clear advantage of the ease of interpretation of the marginal effect. We have re-run all the models using logit and obtained similar results (available upon request) in terms of main significant factors. Table 2 reports the estimated LPM models. The model fit is good overall given a binary nature of the dependent variables. Adjusted R-squared ranges from the minimum of 3.6% for the dependent variable "Fears to lose job" (which is probably not surprising given that fears are often irrational) to 21.7% in the variable "Working from home".

For each dependent variable, let us briefly discuss the key correlates, which are significant at 5%. Model (I) reports that respondents age 25-34 were more likely to be not working by 9.6% points (significant at 5%). Similarly, households with 2 or more members were 11.9% to 17.7% more likely not to work after the beginning of the quarantine compared to single-member households, perhaps, because of cross-insurance available in bigger households. Respondents in certain economic sectors were more likely to remain employed even during the quarantine: by 18.3% points (significant at 5%) in Public administration, by 23.8% points (significant at 1%) in Research, and by 21.6% points in Programming and IT (significant at 1%). Similarly, for certain occupations, employment prospects remain bright even during the epidemic crisis. Highly qualified specialists such as physicians, lawyers and similar are less likely to lose a job by 15.4% while middle and department managers' chances to stay employed are higher by 20.2% and 15.8% points (all coefficients significant at 5%). The probability of staying employed during the quarantine for military and police personnel is higher by 35.9% points compared to the base category of other occupations.

Model (II) reports that working from home part or full time was less likely by 10.6-14.1% points (significant at 5%) for multi-person households compared to single-person households, perhaps, because of more limited working space. On the other hand, females were more likely to work remotely by 12.7% points while better-educated respondents with post-secondary and higher education had chances of working from home higher by 13.8% points (significant at 5%) and 11.8% points (significant at 1%) correspondingly. Living in the capital city of Kyiv was associated with 20.4% points higher probability of remote work compared to the central region with no statistically significant effect for all other regions. Respondents in the sectors of Research and Education were more likely to work from home by 22.3% and 32.7% points correspondingly (both significant at 1%). Finally, highly qualified specialists had 25.6% points (significant at 1%) higher chances of working remotely while middle managers and department managers had chances higher by 33.6% points (significant at 1%) and 15.5% points (significant at 5%), respectively.

"Fears to lose job" model in column (III) is characterized by the lowest goodness-of-fit measure, probably, because fears could be irrational and hard to model based on observable characteristics. Multi-member households are once again very different from a single-member household. Respondents in such households were 14.9-20.5% points more likely to be afraid of losing jobs, perhaps, because of the feeling of responsibility for the other cohabitants and the need for cross-insurance within the family. Interestingly enough, households with children were not different in statistical sense from households without children. Employees in government-related sectors of Public administration, Health care, and Education were less likely to fear job loss by 26.9%, 20.3%, and 26.7% points compared to other sectors. Unlike the previous two models, middle managers felt more insecure and reported a 26.0% higher probability of fear of losing a job.

Finally, we identified the correlates of higher financial insecurity of families with savings sufficient for 1 month or less in Model (IV) for the same sample as in the previous three models for comparison. Besides, we report Model (V) for an extended sample, including all respondents, even those who did not work before COVID-19 but excluding job-related questions. Surprisingly enough, this is the only model with no effect of the household size, which deserves additional investigation given that multi-member households spend less per person compared to a single-member household (Deaton and Paxson 1998). Interestingly enough, knowledge of the actual number of COVID cases in the world was associated with 7.1-9.9% points lower probability of having savings for 1 month or less. Respondents who were more closely following the news about the seriousness of the epidemics could plausibly make additional savings. Unlike previous models, significant differences exist in Model (V) at the regional level – living in the East and South were associated with 10.8% and 6.9% points higher probability of having limited savings compared to the central region (both significant at 5%). Another striking result is that working before the quarantine was not associated with increased financial security in a statistical sense. A phenomenon of "the working poor" is especially pronounced in transition countries like Ukraine, where the monthly minimum wage is below 200 USD.[6]

Table 2. Results of the linear probability model estimation

|  | Not working | Working from home | Fears to lose job | Savings for <1 month | Savings for <1 month |
|---|---|---|---|---|---|
|  | (I) | (II) | (III) | (IV) | (V) |
| Female | -0.062 | 0.127*** | -0.043 | -0.001 | 0.008 |
|  | (0.040) | (0.030) | (0.035) | (0.040) | (0.023) |
| Age 25-34 | 0.096** | -0.097* | 0.090* | 0.078 | 0.053 |
|  | (0.044) | (0.049) | (0.054) | (0.067) | (0.045) |
| Married | -0.099** | 0.043 | -0.069 | -0.109** | -0.089** |
|  | (0.039) | (0.046) | (0.047) | (0.051) | (0.038) |
| Never Married | -0.039 | -0.045 | 0.012 | -0.108* | -0.112** |
|  | (0.059) | (0.063) | (0.066) | (0.063) | (0.047) |
| 2 members in HH | 0.119*** | -0.106** | 0.149*** | 0.002 | 0.024 |
|  | (0.039) | (0.049) | (0.048) | (0.105) | (0.072) |
| 3 members in HH | 0.153*** | -0.116** | 0.198*** | 0.007 | 0.028 |
|  | (0.052) | (0.058) | (0.055) | (0.110) | (0.082) |
| 4+ members in HH | 0.177*** | -0.141** | 0.205*** | 0.071 | 0.061 |
|  | (0.041) | (0.054) | (0.054) | (0.110) | (0.079) |
| Post-secondary education | 0.033 | 0.138** | 0.037 | -0.030 | -0.076 |
|  | (0.067) | (0.054) | (0.114) | (0.068) | (0.060) |
| Higher education | -0.026 | 0.118*** | 0.026 | -0.104* | -0.040 |
|  | (0.058) | (0.037) | (0.072) | (0.058) | (0.045) |

---

[6] https://www.president.gov.ua/en/news/na-vikonannya-zavdannya-prezidenta-uryad-viznachiv-grafik-pi-61929

| | | | | | |
|---|---|---|---|---|---|
| Knows # of COVID cases | -0.052* | 0.090** | -0.028 | -0.071** | -0.099*** |
| | (0.031) | (0.034) | (0.032) | (0.028) | (0.023) |
| East | 0.083* | 0.062 | 0.072* | 0.057 | 0.108** |
| | (0.048) | (0.045) | (0.037) | (0.049) | (0.046) |
| Kyiv | 0.014 | 0.204*** | -0.018 | 0.066 | 0.044 |
| | (0.029) | (0.043) | (0.040) | (0.043) | (0.027) |
| South | 0.047 | 0.090 | 0.022 | 0.049 | 0.069** |
| | (0.055) | (0.078) | (0.038) | (0.043) | (0.029) |
| Middle financial status | 0.004 | 0.011 | -0.045 | -0.133*** | -0.163*** |
| | (0.045) | (0.036) | (0.039) | (0.043) | (0.038) |
| Wealthy financial status | -0.056 | 0.039 | -0.079* | -0.340*** | -0.337*** |
| | (0.048) | (0.039) | (0.040) | (0.045) | (0.027) |
| Public administration | -0.183** | -0.080 | -0.269*** | -0.096 | |
| | (0.079) | (0.074) | (0.072) | (0.102) | |
| Health care and social assistance | -0.086 | -0.190* | -0.203** | -0.133 | |
| | (0.124) | (0.098) | (0.078) | (0.084) | |
| Research | -0.238*** | 0.223*** | 0.057 | -0.277** | |
| | (0.066) | (0.061) | (0.069) | (0.111) | |
| Education | -0.134 | 0.327*** | -0.267*** | 0.003 | |
| | (0.087) | (0.120) | (0.080) | (0.077) | |
| Programming and IT | -0.216*** | 0.111 | -0.088 | -0.196** | |
| | (0.078) | (0.101) | (0.118) | (0.078) | |
| Advertising and mass media | -0.169* | 0.121* | -0.034 | -0.171*** | |
| | (0.086) | (0.070) | (0.088) | (0.062) | |
| Finance, banking and legal | -0.123 | -0.036 | -0.129** | -0.177* | |
| | (0.109) | (0.067) | (0.058) | (0.093) | |
| Owner/co-owner of a large business | 0.223 | 0.134 | 0.976*** | 0.116 | |
| | (0.382) | (0.326) | (0.116) | (0.267) | |
| Army and police | -0.359*** | -0.003 | -0.072 | -0.021 | |
| | (0.103) | (0.201) | (0.117) | (0.183) | |
| Highly qualified specialist | -0.154** | 0.256*** | 0.142* | 0.138 | |
| | (0.072) | (0.067) | (0.074) | (0.094) | |
| Middle manager | -0.202** | 0.336*** | 0.260*** | 0.069 | |
| | (0.096) | (0.093) | (0.087) | (0.097) | |
| Department manager | -0.158** | 0.155** | 0.119 | 0.078 | |
| | (0.077) | (0.066) | (0.078) | (0.091) | |
| Constant | 0.292 | 0.200 | 0.086 | 0.839*** | 0.730*** |
| | (0.191) | (0.161) | (0.183) | (0.146) | (0.083) |
| Observations | 800 | 800 | 800 | 800 | 1147 |
| Adjusted R-squared | 0.096 | 0.217 | 0.036 | 0.123 | 0.111 |

*Notes: Authors' calculations based on the Gradus survey. All models use robust standard errors clustered at the city level. Model control for all explanatory variables, but only coefficients significant at 5% or less are reported to save space. ***$p < 0.01$; **$p < 0.05$; *$p < 0.1$.*

## 5 Discussion and conclusions

In this paper, we report the evidence on job market outcomes and expectations of urban Ukrainian households four weeks after the beginning of the quarantine. Using the unique dataset which is nationally representative of the urban adult population in Ukraine, we provide important insights into the effects of the COVID imposed job restrictions on employment and well-being.

Our findings indicate potential concerns for the well-being of urban Ukrainians. The crisis is likely to exacerbate socio-economic inequalities: better educated respondents and those living in Kyiv are more likely to secure work from home. In terms of policy suggestions, the government of Ukraine should pay more attention to the labor market in regions and provide social support to lower-skilled social groups.

Some important socio-demographic differences also exist. Women, who typically take care of household tasks, are also likely to combine these tasks with the role of a breadwinner by working from home. The analyses also show the role played by the household size – respondents from bigger households are less likely to secure their jobs during the quarantine and to work from home and also are more likely to be afraid of the job loss.

Another set of findings may be unique to a transition country like Ukraine. For example, better educated respondents are not protected from job loss or financial insecurity, which is also not improved if a respondent was employed before the quarantine. These findings may indicate low quality of education and a phenomenon of a "working poor" still characterizing Ukraine.

While we do realize that our results do not have causal interpretation, they still identify key correlates of the short-term job market outcomes as well as expectations regarding the new quarantine reality.

# Appendix

Table 3. Means and standard deviations of employment sectors and occupations

| Variables | # of Obs. | Mean | St. Dev. |
|---|---|---|---|
| Public administration | 874 | 0.042 | 0.201 |
| NGO | 874 | 0.022 | 0.146 |
| Health care and social assistance | 874 | 0.064 | 0.245 |
| Culture, sports and entertainment | 874 | 0.039 | 0.193 |
| Research | 874 | 0.047 | 0.212 |
| Education | 874 | 0.093 | 0.290 |
| Hotels and restaurants | 874 | 0.024 | 0.153 |
| Programming and IT | 874 | 0.079 | 0.270 |
| Manufacturing | 874 | 0.125 | 0.331 |
| Advertising and mass media | 874 | 0.062 | 0.241 |
| Agriculture, forestry and fishing | 874 | 0.022 | 0.146 |
| Construction | 874 | 0.038 | 0.191 |
| Sales | 874 | 0.142 | 0.349 |
| Transport and communications | 874 | 0.054 | 0.226 |
| Finance, banking and legal | 874 | 0.059 | 0.237 |
| Owner/co-owner of a large business | 924 | 0.002 | 0.046 |
| Owner/co-owner of medium/small business | 924 | 0.031 | 0.174 |
| Army and police | 924 | 0.014 | 0.118 |
| Highly qualified specialist | 924 | 0.216 | 0.412 |
| Director | 924 | 0.013 | 0.113 |
| Skilled worker | 924 | 0.179 | 0.383 |
| Middle manager | 924 | 0.140 | 0.347 |
| Unskilled worker | 924 | 0.036 | 0.186 |
| Artistic and creative occupations | 924 | 0.037 | 0.188 |
| Department Manager | 924 | 0.118 | 0.323 |
| Clerk | 924 | 0.085 | 0.280 |
| Self employed | 924 | 0.085 | 0.280 |

*Notes: Authors' calculations based on the Gradus survey.*